\documentclass[onecolumn,aps,prr,preprintnumbers,amsmath,amssymb]{revtex4}

\usepackage{natbib}
\usepackage{graphicx}
\usepackage{bm}
\usepackage{color}

\usepackage{amsmath}
\usepackage{amsfonts}
\usepackage{amssymb}
\usepackage{hyperref}
\usepackage{makeidx}
\usepackage{graphicx}
\begin{document}
	
	\title{Dephasing effects on quantum correlations and teleportation in presence of state dependent bath}
	\author{Mehboob Rashid$^1$}
	\author{Muzaffar Qadir Lone \footnote{corresponding author: lone.muzaffar@uok.edu.in}$^2$}
	\author{Prince A Ganai$^1$}
	\affiliation{$^1$Department of Physics, National Institute of Technology, Srinagar-190006 India.\\
		$^2$Quantum Dynamics Lab, Department of Physics, University of Kashmir, Srinagar-190006 India}
	
	\begin{abstract}
		Quantum information protocols are often designed in the ideal situation with no decoherence. However, in real setup, these protocols are subject to the decoherence and thus reducing  fidelity of the measurement outcome. In this work, we analyze the effect of state dependent bath on the quantum correlations  and the  fidelity of a single qubit teleportation. We model our system-bath interaction  as  qubits interacting with  a common bath of bosons, and the state dependence of the bath is generated through a projective measurement on the joint state in thermal equilibrium. The analytic expressions for the time evolution of entanglement, discord and average fidelity of quantum teleportation are calculated. It is shown that due to the presence of initial system-bath  correlations, the system maintains quantum correlations for long times. Furthermore, due to the presence of finite long time entanglement of the quantum channel, the average fidelity  is shown to be higher than its classical value. 
		\end{abstract}
	\maketitle
	\section{Introduction}

Quantum correlations described by entanglement\cite{1}  and discord\cite{2,3,4,5,6} are important features of quantum mechanics that arise due to non-separbility, non-locality or impossibility of local descrimination. In addition to  their role in fundamental aspects of physics, these correlations find their applications as the resource for quantum computation and quantum information\cite{7,8}. 
For example, quantum teleportation\cite{9,10,11,12}, super dense coding\cite{13,14,15,16}, etc. In many ways, quantum communication protocols are superior to their conventional counterparts\cite{17,18,19,20,21,22}. For instance, they feature excellent security and channel capacity\cite{23,24,25}. Quantum teleportation protocol is one of the several techniques that allow for a  unit fidelity 
of a quantum state transfer with preshared maximal entangled state between two parties.  Furthermore, certian class of separable states with non-zero discord have been recognised as resource for speed up of certian computational tasks over classical counterparts\cite{26,27,28,29,30}.

In contrast to isolated quantum systems, the interaction of a system with the bath degrades quantum correlations\cite{31}. This in turn affects the utilization of quantum correlations for quantum technologies. 
 The effects of these system-bath (SB) interactions lead to Markovian or non-Markovian dynamics. In Markovian case, the dynamics is memoreyless, while in non-Markovian dynamics, the system retrieves information back from the bath signalling the presence of memory effects\cite{32,33}. The non-Markovian effects have shown to play a significant role for various quantum protocols like the dissipative quantum computation\cite{34,35}, quantum metrology\cite{36,37,38,39}, entanglement generation\cite{40,41,42}, dynamical control of correlations in various system like quantum optics\cite{43,44,45,46}, nuclear magnetic resonance \cite{47},
nanophysics\cite{48,49,50}, etc. In understanding such dynamics, it is often assumed that system and bath are initially uncorrelated which is a consequence of Born approximation. However, under strong coupling, this assumption is violated, for example in quantum state preparation. In this direction many works have analyzed the role of these initial SB correlations in dephasing models\cite{51,52,53,54,55,56}, superconducting qubits\cite{57}, quantum dots\cite{58} etc.

In this work, we consider a dephasing model represented by two qubits coupled to a collective bath  with distance dependent interaction. Our goal is to study the role of initial SB interactions on the dynamics of quantum correlations and quantum teleportation. In  earlier works, effects of such initial SB correlations have been studied. For example,  Li, et. al.\cite{59} Zhang, et. al. \cite{60}
 have shown that initial SB correlations have strong influence on the dynamics of quantum discord and entanglement. However, the type of initial states considered in these works are restricted to pure states at zero temperature only. Here we consider a class of initial states at finite temperature obtained via projective measurements.
Furthermore, the dynamics of average fidelity of teleportation of a single qubit in presence of some particular types of noise have also been studied\cite{61,62,63,64,65}.
It has been shown that local noise can even boost the fidelity of single-qubit teleportation \cite{66,67,68,69}. In these works, initial SB correlations present in the teleportated qubit or in the entangled channel are not considered. Here we attempt to  analyze whether these initial SB correlations affect the average fidelity of teleportation.

This paper is structured in the following way. We introduce the model system with SB correlations in section II.  
The dynamics of quantum correlations given by negativity and discord is presented in section III. In section IV, we discuss quantum teleporation protocol and find that the initial SB correlations help to maintain average fidelity above classical value. Finally we conclude in section V.

\section{Modelling system-bath interactions}
We consider a two qubit channel shared by Alice and Bob that evolves according to a  dephasing model where each qubit separated by distance $L$, is coupled to collective bath  given by  ($\hbar = 1$): 
 	\begin{eqnarray}
 		\label{6}
 	H &=& H_S + H_B + H_{int} \nonumber \\
 	&=&\frac{\omega_0}{2}\sum_i\sigma_{i}^z + \sum_{k}\omega_{k}b_{k}^{\dagger}b_k + \sum_{ik}\sigma_{i}^z(g_{k}e^{-i\vec{k}.\vec{r_i}}b_k + h.c.) .
 \end{eqnarray}
 $\omega_0$ is energy splitting of qubits; bath modes are characterized by energy  $\omega_k$ with $b_k,b_k^{\dagger} $ as annihilation and creation operators for the  $k$th bath mode. $\sigma^{z}_i$ and  $\vec{r_i}$ are the $z$-Pauli matrix and  position vector of $ith$-qubit respectively. Here, h.c. means Hermitian conjugate. For notational convenience  we call channel shared by Alice and Bob as ``system (S) ".  Since in the Born approximation, system-bath correlations are neglected. However, in this work we consider a particular type of initial state that incorporates system-bath correlations at finite temperature\cite{70,71}. In order to generate a state dependent bath i.e. initial system bath correlations, 
we consider a thermal equillbrium state given by $	\rho_{SB}^T  = \frac{e^{-\beta H }}{Z}$. Here $ Z $ is the partition function and $\beta =\frac{1}{T}$. Now we make a projective measurement via  projection operators $\{\Pi_i\}$ on the state of the system such that the total SB density operator collapses to 
 
 \begin{eqnarray}
     \rho_{SB}^T = \frac{1}{Z}\sum_i \Pi_i e^{-\beta H }\Pi_i.
 \end{eqnarray}
 Now we prepare the state of the system to be in
 $|\psi\rangle$ so that $\Pi_i=|\psi\rangle \langle \psi|$. With this projection, the above sum reduces to a single term:
\begin{eqnarray}
\label{cors}
    \rho_{SB}^T =|\psi\rangle  \langle \psi| \otimes \frac{1}{Z_B}   \langle \psi|     e^{-\beta H }    |\psi\rangle  = \rho_S \otimes \rho_B^{\psi}
\end{eqnarray}
 where $Z_B = Tr_{B}\langle{\psi}|\exp(-\beta H)|{\psi}\rangle$. { First, we compare this state with the uncorrelated state used via Born approximation: $ \rho_{tot}(0)= \rho_S (0) \otimes  \rho_B$ where $\rho_B = \frac{e^{-\beta H_B }}{Z} $ is the bath density matrix. Here the bath state does not depend on the parameters of the system while the bath state defined in equation \ref{cors} depends non-trivially on the paramters of state of the system $|\psi\rangle$. Next, we make comparison to the correlated initial states reported in the literature for example in references \cite{59,60}, which 
are of the form $ 	|\psi\rangle_r =\alpha |0\rangle_S|0\rangle_B + \beta |1\rangle_S|1\rangle_B$, 
where $|0\rangle_B$ is the vacuum state of the bath and $|1\rangle_B =  b^\dagger |0\rangle$ is a bath state with single excitation. This assumption is adhoc in a sense that it does not incorporate more number of excitations which are important in strong coupling limit or when Born approximation is not valid. These are pure states with no temperature dependence. The form of initially correlated states considered in the equation \ref{cors} are entirely different in their construction. These states arise due to selective measurements on a pre-defined thermal equilibrium state which give rise to non-trivial initial system-bath correlations. In contrast, the measurement on the state $|\psi\rangle_r$ results in an uncorrelated state. Furthermore, a pecularity of the state in equation \ref{cors} is its variation of the SB correlation with temperature. As  $T \rightarrow 0$, the joint state of the system and bath $\rho = e^{-\beta H}/{Z} \rightarrow |{\rm gnd}\rangle\langle {\rm gnd}|$, where $|{\rm  gnd}\rangle$ is some ground state of the Hamiltonian system (S) plus bath (B) $H_S$ +$H_B$ becomes uncorrelated whereas for T$\ne$ 0, the joint state becomes correlated due to thermal fluctuations. The vice-versa happens for $|\psi\rangle_r$. The total state in equation \ref{cors} is a non-Gibb's state.

 }

  Next, we consider the initial state of the system to be $|\psi\rangle_S= \sqrt{\alpha}|00\rangle + \sqrt{1-\alpha}|11\rangle$; therefore, we can calculate density matrix  of the bath as:
  \begin{eqnarray}
		\rho_B^{\psi} (0)&=& \frac{1}{Z_B}   \langle \psi|     e^{-\beta H }    |\psi\rangle \nonumber \\
  &=&\frac{ 1}{Z_B}\big[ \alpha e^{-\beta\omega_0}e^{-\beta H_{1}^{+}}+ (1-\alpha) e^{\beta\omega_0} e^{-\beta H_{1}^{-}})  \big], 
	\end{eqnarray}
where $ H_{1}^{\pm}  = H_B \mp (B_{1k} \pm B_{2k})$ and $B_{ik}= g_{ik}b_k+ g_{ik}^{\star}b_k^{\dagger}$, $i=1,2$. Now, we  consider the time evolution operator $U(t) =Te^{-i \int_0^t d\tau H_I(\tau)}$, $T$ is the time ordering operator $H_I(t)$ is the interaction Hamiltonian in interaction picture. The time evolved density matrix of the system can be calculated using the depahsing model above and is given by \cite{71}
\begin{eqnarray}
 	\label{8}
 	\rho_{S}(t) &=& {\rm Tr}_B[U(t) \rho^T_{SB}(0) U(t)^{\dagger}]={\rm Tr}_B [U(t) \rho_S \otimes \rho_B^{\psi} U(t)^{\dagger} ]  \nonumber \\
 &=& \alpha |00\rangle \langle 00| + \sqrt{\alpha(1-\alpha)} \kappa(t)|00\rangle \langle 11| + 
 \sqrt{\alpha(1-\alpha)} \kappa^*(t) |11\rangle \langle 00| + (1-\alpha) |11\rangle \langle 11|
  \end{eqnarray} 
where 
\begin{eqnarray}
\kappa(t)	=	\Bigg[ \frac{\alpha e^{-\beta\omega_0} e^{2i\zeta(t)}+(1-\alpha) e^{\beta\omega_0} e^{-2i\zeta(t)} }{\alpha e^{-\beta\omega_0} +(1-\alpha) e^{\beta\omega_0})} \Bigg]\exp(-4\Sigma_{k}\frac{|g_k|^2}{\omega_{k}^2}\big(1+\cos{\Vec{k}.\Vec{L}}\big)(1-\cos{\omega_{k}t}) \coth{\frac{\beta\omega_{k}}{2}})
\end{eqnarray}
 and $\zeta(t)= 8 \sum_k \frac{|g_k|^2}{\omega_k^2} \sin \omega_k t \Big[1+\cos (\vec{k}.\vec{L})\Big] \nonumber$. The  term in square brackets of $\kappa(t)$ capture the initial SB correlations with dependence on the system parameters, while exponential is the standard decoherence function with dependence on distance of separation of qubits in addition to the dependence on  temperature and bath parameters.  To simplify analysis, we use $\alpha=1/2$.

 \section{Dynamics of Quantum Correlations}
 \subsection{Entanglement}
 In this section, we study the time evolution of entanglement of the two-qubit quantum channel represented by the aforementioned model via negativity($\mathcal{N}$). This measure os based on positive partial transpose (PPT) criteria for the separability
  and is defined as\cite{72} :
 \begin{eqnarray}
 	\mathcal{N(\rho)} := \frac{|| \rho^{T_A}||-1}{2}
 \end{eqnarray}
where, $\rho^{T_a}$ is the partial transpose of $\rho$ with respect to subsystem A. The trace norm of an operator $\hat{O}$ is given as
$||\hat{O}||_1=Tr|\hat{O}|$=Tr$\sqrt{O^\dagger O}$. For the two qubit channel given by the class of states  $|\psi\rangle_S=\sqrt{\alpha}|00\rangle + \sqrt{1-\alpha}|11\rangle $, with $0\le \alpha\le 1$, we can write entanglement negativity as
  \begin{eqnarray}
 	\mathcal{N}(\rho)= 2\sqrt{\alpha(1-\alpha)} \sqrt{[\cos^2(\zeta(t)) + \sin^2(\zeta(t)) \tanh^2(\beta \omega_0)}] e^{-\gamma_s(t)}.
 \end{eqnarray}
% \begin{eqnarray}
 %	\mathcal{C}(t)= {\rm max}\{0, 2\sqrt{p(1-p)} \sqrt{[\cos^2(\zeta(t)) + \sin^2(\zeta(t)) \tanh^2(\beta \omega_0)}] e^{-\gamma_s(t)}\}.
% \end{eqnarray}
 %The square root factor in the above formula are due to initial SB correlations while $\gamma_s(t)$ is the standard decoherence function. In other words, we write time evolved concurrence as 
% $ 	\mathcal{C}(t)= e^{-\Gamma(t)} C(0)$
%  with $\Gamma(t) =\gamma_{ic}(t)+ \gamma_s (t) $ and $C(0)=2\sqrt{p(1-p)} $.
   The standard decoherence function $\gamma_s(t)$ is given by 
 \begin{eqnarray}\label{9}
 	\gamma_s (t)= 4\sum_k |g_k|^2 \frac{\cos^2(\vec{k}.\vec{L})}{2} \frac{1-\cos \omega_k t}{\omega_k^2} \coth \frac{\beta \omega_k}{2} 
 \end{eqnarray}
 while the decoherence due to initial SB correlations are encoded in the function $\gamma_{ic}(t)$:
 \begin{eqnarray}
 	\gamma_{ic}(t) =-\frac{1}{2} \log [ \cos^2(\zeta) + \sin^2(\zeta) \tanh^2(\beta\omega_0) ].
 \end{eqnarray}

 \subsection{Quantum Discord}
 Quantum discord represents quantum correlations beyond entanglement. It is defined as the difference between total correlations and classical correlations in a given system.  Let $\rho^{AB}$ be the density operator for a bipartite system AB, then the total correlations are given by the mutual information $I(\rho^{AB})$:
 \begin{eqnarray}
 	I(\rho^{AB}) = S(\rho^A) + S(\rho^B)-S(\rho^{AB})
 \end{eqnarray}
 where $S(\rho)=-tr (\rho log \rho)$ is the von Neuman entropy of the density matrix $\rho$. In order to determine the classical correlations, we define one dimensional projectors ${P_k}$, so that the conditional density matrix after measurements on the subsystem B, can be written as
 $\rho_k =\frac{1}{p_k} (I_A \otimes P_k) \rho^{AB} (I_A\otimes P_k)$, with $p_k = tr((I_A \otimes P_k) \rho^{AB} (I_A\otimes P_k)) $. Therefore, we write the entropy corresponding to this measurement as $S(\rho^{AB}|P_k) = \sum_k p_k S(\rho_k)$. The mutual information after this measurement can be written as
 $I(\rho^{AB}|P_k)= S(\rho^{A})-S(\rho^{AB}|P_k) $. Therefore, we write the classical correlations $C(\rho^{AB})$ present in the quantum system as the supremum over all von Neuman measurements ${P_k}$:
 \begin{eqnarray}
 	C(\rho^{AB})= \underset{\{P_k\}}{{\rm sup}}~I(\rho^{AB}|P_k).
 \end{eqnarray}
 \begin{figure}[t]
 	\includegraphics[width=4 cm,height=4cm]{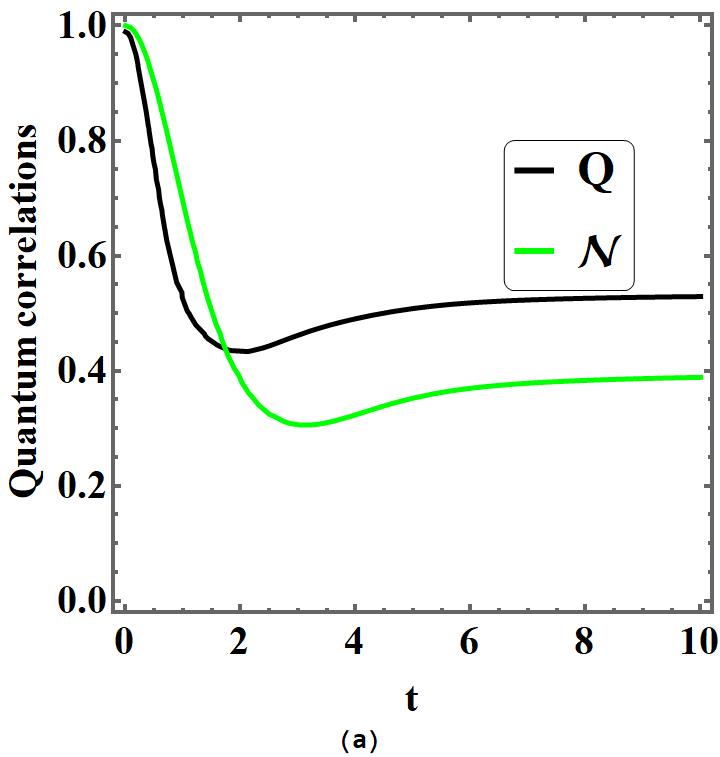} \hspace{2mm}
 	\includegraphics[width=4cm,height=4cm]{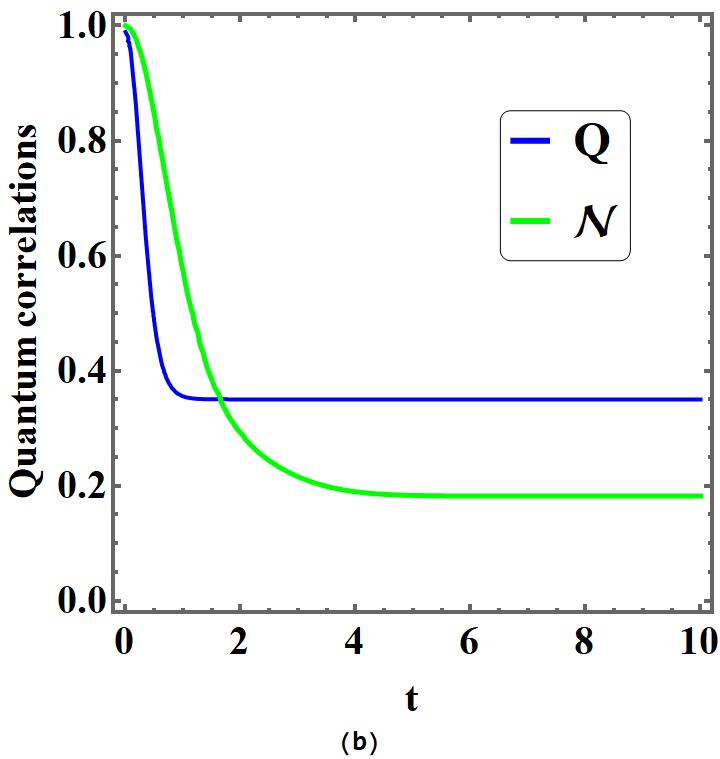}\hspace{2mm}
 	\includegraphics[width=4cm,height=4cm]{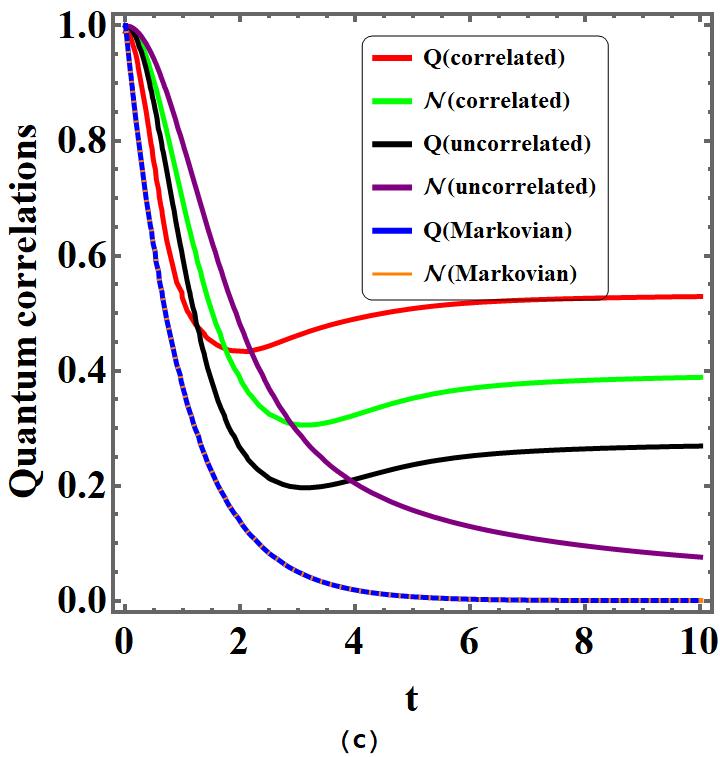} \hspace{2mm}
 	\includegraphics[width=4cm,height=4cm]{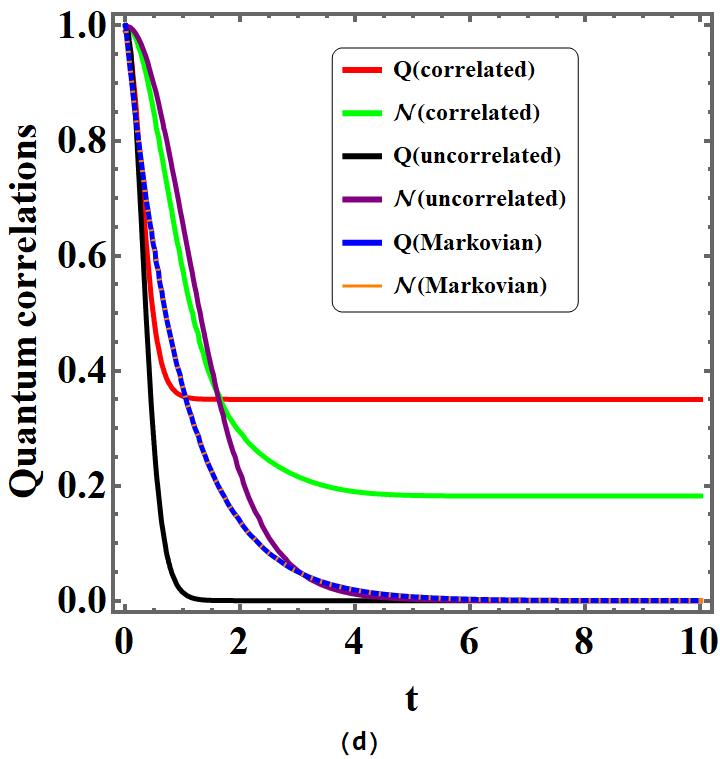} \hspace{2mm}
 	\caption{Variation of quantum discord $Q$ and  negativity $\mathcal{N}$ with respect to time $t$ for temperatures (a) $\beta \omega_c \sim 1$ (b)$ \beta\omega_c << 1$. Here we have taken $s=1$. A comparison is made with respect to uncorrelated initial SB state and that of Markovian case for the same temperature ranges (c) $\beta \omega_c \sim 1 $, (d) $ \beta\omega_c << 1$. We see that non-Markovian evolution in presence of initial SB correlations help to maintain non-zero negativity and discord for long times.}
 	\label{fig1}
 \end{figure}
 Quantum discord $Q(\rho^{AB})$ is therefore given as the difference between mutual information $I(\rho^{AB})$ and the classical correlations $C(\rho^{AB})$\cite{2,5}:
 \begin{eqnarray}
 	Q(\rho^{AB})= I(\rho^{AB})-C(\rho^{AB}).
 \end{eqnarray}
 This expression is in general difficult to evaluate. However, the initial states considered here we can calculate it exactly and is given by \cite{73}
 \begin{eqnarray}
 	Q= {\rm min}(Q_1, Q_2)
 \end{eqnarray}
 with $Q_1=1$ and 
 \begin{eqnarray}
 	Q=  \Big(\frac{1+ |\kappa(t)|}{2}\Big)\log_2\Big(\frac{1 + |\kappa(t)|}{2}\Big) + \Big(\frac{1 - |\kappa(t)|}{2}\Big)\log_2\Big(\frac{1 - |\kappa(t)|}{2}\Big) +1    
 \end{eqnarray}
 
 Next, in order to analyze the behaviour of quantum correlations given by concurrence and quantum discord, we first define bath spectral density $J(\omega)$ as
 \begin{eqnarray}
 	J(\omega) = \sum_k |g_{1k}+ g_{2k}|^2 \delta(\omega-\omega_k).
 \end{eqnarray}
 The exact form of this function is very complicated and depends on dimensionality of the bath, however we can model it phenomenologically. We assume the form of $g_k$ in $\omega$-space as $g(\omega)=\eta \frac{\omega}{\omega_c} e^{-\frac{\omega^2}{\omega^2_c}}$, where $\eta $ is the intrinsic SB coupling
 and $\omega_c$ is the cutoff frequency of the bath. Using this form of $g(\omega)$ and integrating over solid angle in equations (for gamma and phi)
 we get
 \begin{eqnarray}
 	\gamma_s(t) &=& \frac{8 \eta}{ \omega_c^2 \pi^2 c^3} \int_0^{\infty} d\omega \omega e^{-\frac{\omega^2}{\omega^2_c}}\Big(1+\frac{\sin \omega s}{\omega s}\Big) \sin^2\frac{\omega t}{2}\coth\frac{\beta \omega}{2} . \\
 	\zeta(t) &=& \frac{4\eta }{\pi^2 c^3\omega_c^2} \int_0^{\infty} d\omega \omega e^{-\frac{\omega^2}{\omega_c^2}}\Big(1+\frac{\sin \omega s}{\omega s}\Big) \sin \omega t .
 \end{eqnarray}
 Here $s=\frac{L}{c}$ with $c$ to be the velocity of the bath modes and the $L$ the distance of qubits of the channel. $s$ defines a time scale due to the SB interaction mediated by bath modes. 
 Furthermore, different time scales that arise in our model are provided by the cutoff frequency $\omega_c$ which give the relaxation time scale for the bath; relaxation time scale for qubits provided by energy $\omega_0$.  Using these time scales, we parameterize above equations as: $\omega \to \frac{\omega}{\omega_c},~t \to \omega_c t, s\to \omega_c s$ and measure temperature with respect to $\omega_c$: $ \beta \to \beta \omega_c$. For notational convenience,  we take $\omega_c=1$ without loss of generality.
 
 %%%%
 From the expression in equation[\ref{9}], we see that for $\cos(\vec{k}.\vec{L})=-1$, both $\gamma(t) $ and $\zeta(t)$ vanish, thus resulting in no decay of negativity. Since, discord depends on the $|\kappa(t)|$, which in this case turns out to be  $|\kappa(t)|= \frac{1}{\sqrt{2}}\frac{\sqrt{\cosh2\beta \omega_0}}{\cosh\beta \omega_0}$ . Therefore, even though initial correlations does not have any effect on the entanglement decay but strongly effects the discord.  Next, in figure \ref{fig1}, we plot negativity $\mathcal{N}$ and discord $Q$ with respect to rescaled time $t$ for $\beta \omega_c \sim 1$ and $\beta \omega_c << 1$, which correspond to low and high temperatures. In the low temperature regime $\beta \omega_c \sim 1$ (fig. \ref{fig1}(a)), we observe that both negativity $\mathcal{N}(t)$  and discord  $Q(t)$ have  non-monotonic behaviour, that decays initially but finally saturates to a finite non-zero value. However, at high temperature $\beta \omega_c <<1$ (fig. \ref{fig1}(b)), we see that quantum discord initially has an abrupt decay in comparison to negativity with a saturation to a non-zero value in the long time limit. We can understand this behavior in an intuitive way as follows. The number of modes that would cause fast decoherence are suppressed by the initial SB correlations. However, at high temperatures, we have thermal fluctuations which increases the number of modes to be scattered causing an abrupt decay of correlations; the competition between these thermal fluctuations with those of initial SB correlations result in a less finite value in comparison to its low temperature case. This can be further verified from the comparison to the Markovian and uncorrelated initial SB states plotted  in figure \ref{fig1}(c) and \ref{fig1}(d) for low and high temperature regimes respectively. We see that initial SB correlations help to maintain coherence in the system  for long times.

\section{Quantum Teleportation in presence of dephasing}
\subsection{ Standard Teleportation Protocol}
In standard quantum teleportation protocol\cite{9}, an unknown quantum state $ |\psi_{int}\rangle $ is teleported from Alice to Bob who share an entangled state that acts as a quantum channel between them. The protocol can be formulated in terms of density matrix formalism as follows. Let $\rho_{in}$ be the density matrix of an unknown state to be teleported, $\rho_{AB}$ be the channel density matrix shared by Alice and Bob;  and $\rho_{B}$ be the output density matrix i.e. density matrix of the teleported state recovered by  the Bob. The total initial state ($\rho_{in}$ and the channel) is given by
\begin{eqnarray}
\rho^T = \rho_{in} \otimes \rho_{AB}.
\end{eqnarray}
As the first step of the protocol, Alice performs projective measurements on her qubits, namely the input state and her portion of the entangled channel. Let $\{\Pi_i\}$ be the set of projection operators used by Alice. Thus after the projective measurements, the state of the total system changes to 
\begin{eqnarray}
	\label{2}
 \rho_i^T = \frac{\Pi_i \rho^T \Pi_i}{P_i} 
\end{eqnarray}
where $P_i={\rm Tr}\Pi_i \rho^T \Pi_i $ is the probability of occurrence of specific density matrix $\rho_i^T$ corresponding to $\Pi_i$-projection. As a next step, Alice communicates these measurement results to the Bob via a classical channel. With this knowledge Bob recovers the teleported state $\rho_B$ by applying suitable unitary operators to his density matrix:
\begin{eqnarray}
\rho_i^B = \frac{U_i {\rm Tr}_A [\rho^T_i] U^{\dagger}_i}{Q_i}= \frac{U_i {\rm Tr}_A [\Pi_i \rho^T \Pi_i] U^{\dagger}_i}{Q_i}.
\end{eqnarray}
Here ${\rm Tr}_A $ means trace over the Alice's qubits. The unitary operators $U_i$, which Bob must apply to complete the protocol, is dependent not only on Alice's measurement outcome but also on the quantum channel that was employed. As a specific example, which we consider in this work, is the teleportation of a single qubit  $|\psi\rangle = \cos{\frac{\theta}{2}}|0\rangle + \sin{\frac{\theta}{2}}e^{i\phi}|1\rangle$ ( $\theta$ and $\phi$ are the polar and azimuthal angles) through a noisy channel of two qubits shared by Alice and bob. The $\{\Pi_i\}$ projection operators are specified by Bell states while the unitary operators $U_i$ are given by $\{I, \sigma^x, \sigma^y, \sigma^z\}$ depending on the measurements due to Alice. Here $I$ is the identity operator while $\sigma^i$ $(i=x,~y,~z)$ are the Pauli spin operators.

The performance of a given teleportation protocol can be represented by Fidelity $F$. It represents the overlap of the initial state with the final output state:
\begin{eqnarray}
F= {\rm Tr} [\rho_{in} \rho^B] = \langle \psi_{in}|\rho^B |\psi_{in}\rangle.
\end{eqnarray}
The fidelity is bounded as $0\le F\le 1$, where $F=0$ (no teleportation ) means initial and final states are orthogonal to each other while $F=1$ (perfect teleportation) means initial and final states are same. The classical bound on fidelity is $F=\frac{2}{3}$ which is simulated by classical channel. 
 Since the state to be teleported is typically unknown, it is more practical to determine the average fidelity provided by
\begin{eqnarray}
	\label{5}
		F_{av}= \frac{1}{4\pi}\int_{0}^{\pi} d\theta \int_{0}^{2\pi} d\phi F(\theta,\phi) \sin\theta
\end{eqnarray}
 where 4$\pi$ is the solid angle. 
%\begin{figure}[ht]
   % \includegraphics[width=3.5cm,height=4cm]{LTF.jpg} \hspace{2mm}
    % \includegraphics[width=3.5cm,height=4cm]{HTF.jpg} \hspace{2mm}
     %\includegraphics[width=3.5cm,height=4cm]{cltf1.jpg} \hspace{2mm}
     %\includegraphics[width=3.5cm,height=4cm]{chtf1.jpg}
    %\caption{Variation of quantum discord $Q$ and  Concurrence $C$ with respect to time $t$ for temperatures (a) $\beta \omega_c <<1$ (b)$ \beta\omega_c \sim 1$. Here we have taken $s=1$.}
    %\label{fig2}
%\end{figure}
 \subsection{ Teleportation in presence of initial SB correlations}
Here we consider the following cases to understand the influence of dephasing in presence of initial SB correlations on quantum teleportation. As a first case, we consider the channel shared by Alice and Bob coupled to the bath. We consider the teleportation of a single qubit state $\rho_{in}=|\psi_{in} \rangle \langle \psi_{in} |$, with $|\psi\rangle = \cos{\frac{\theta}{2}}|0\rangle + \sin{\frac{\theta}{2}}e^{i\phi}|1\rangle$. Also, we assume a   two qubit channel (system) shared by  Alice and Bob given by 
  $|\psi\rangle_{S}= \sqrt{\alpha}|00\rangle  + \sqrt{1-\alpha}|11\rangle$ with $0\le \alpha \le 1$. The first qubit is in possession of Alice while Bob holds second qubit. Since the channel is coupled with the bath and the  joint state of the SB evolves according to  the dephasing model given above.  Due to initial SB correlations, the state of bath depends in a non-trivial way on the parameters of the channel and accordingly the  density matrix for the channel can be written as $\rho_S(t)$ given in equation \ref{8} .
Using this channel $\rho_S(t)$, Alice  can teleport a given state $|\psi_{in }\rangle$ faithfully to Bob. To achieve this, Alice performs Bell measurements  on her qubits using projection operators defined by $\{\Pi_i\}$ where 
$\Pi_1 = |\Phi^+\rangle\langle\Phi^+| $, $\Pi_2 = |\Phi^-\rangle\langle\Phi^-| $, $\Pi_3 = |\Psi^+\rangle\langle\Psi^+| $ and  $\Pi_4 = |\Psi^-\rangle\langle\Psi^-| $. The Bell states are defined as $ |\Phi^\pm\rangle=\frac{1}{\sqrt{2}}(|00\rangle\pm|11\rangle)$ and $|\Psi^\pm\rangle=\frac{1}{\sqrt{2}}(|01\rangle\pm|10\rangle)$.
Using the teleportation protocol given above, the state of the Bob (without applying unitary operation) corresponding to projection $\Pi_1$ is given by 

\begin{eqnarray}
\rho^B_1 &=& \frac{1}{4Q_1} \Big[  \alpha \cos^2{\frac{\theta}{2}} |0\rangle \langle 0|  +  \frac{1}{2}\sqrt{\alpha(1-\alpha)} \sin \theta \kappa(t) |0\rangle \langle 1| \nonumber \\
&& + \frac{1}{2}\sqrt{\alpha(1-\alpha)} \sin \theta \kappa^*(t) |1\rangle \langle 0| +  (1-\alpha) \sin^2{\frac{\theta}{2}}  |1\rangle \langle 1|] \Big]
\end{eqnarray}
where  $Q_1 = \frac{1}{2}(1-\alpha \sin^2\frac{\theta}{2}) $. Based on the measurement outcome of Alice, Bob applies now unitary transformation $U=I$ on his qubit to get the output state of the teleportation:
\begin{eqnarray}
    \rho^B_{out_1} =  U_1\rho^B_1 U_1^\dagger =\rho^B_1.
\end{eqnarray}
Next, for the projective measurement by Alice using $\Pi_2$, the state of Bob is given by 
\begin{eqnarray}
\rho^B_2 &=& \frac{1}{4Q_2} \Big[  \alpha \cos^2{\frac{\theta}{2}} |0\rangle \langle 0|  -  \frac{1}{2}\sqrt{\alpha(1-\alpha)} \sin \theta \kappa(t) |0\rangle \langle 1| \nonumber \\
&& - \frac{1}{2}\sqrt{\alpha(1-\alpha)} \sin \theta \kappa^*(t) |1\rangle \langle 0| +  (1-\alpha) \sin^2{\frac{\theta}{2}}  |1\rangle \langle 1|] \Big].
\end{eqnarray}
with $Q_2=Q_1$. Bob now applies unitary transformation $U_2= \sigma^z$ to get the teleported state
\begin{eqnarray}
     \rho^B_{out_2} =  U_2\rho^B_2 U_2^\dagger= \sigma^z \rho^B_2 \sigma^z.
\end{eqnarray}
 Along the similar lines, Bob applies unitary transformation $U_3= \sigma^x,~U_4= \sigma^x\sigma^z$ corresponding to projective measurement of Alice by $\Pi_3,\Pi_4$ to get the teleported state $\rho^B_{out_3},~\rho^B_{out_4}$ respectively with the probabilities $Q_3,~Q_4$. Since different $\rho^B_{out_i}~i=1,2,3,4$
occur in general with different probabilities $Q_i$ so we take average over all $Q_i$ i.e. $\tilde{F}=\sum_i Q_i F_i$. Since, $\tilde{F}$ depends on the input states $|\psi_{in}\rangle$, therefore assuming uniform distribution of all these states we write the efficiency of the protocol in terms of average fidelity $F_{av}$ ($\alpha =1/2$) as:
\begin{figure}[t]
	\includegraphics[width=4cm,height=4cm]{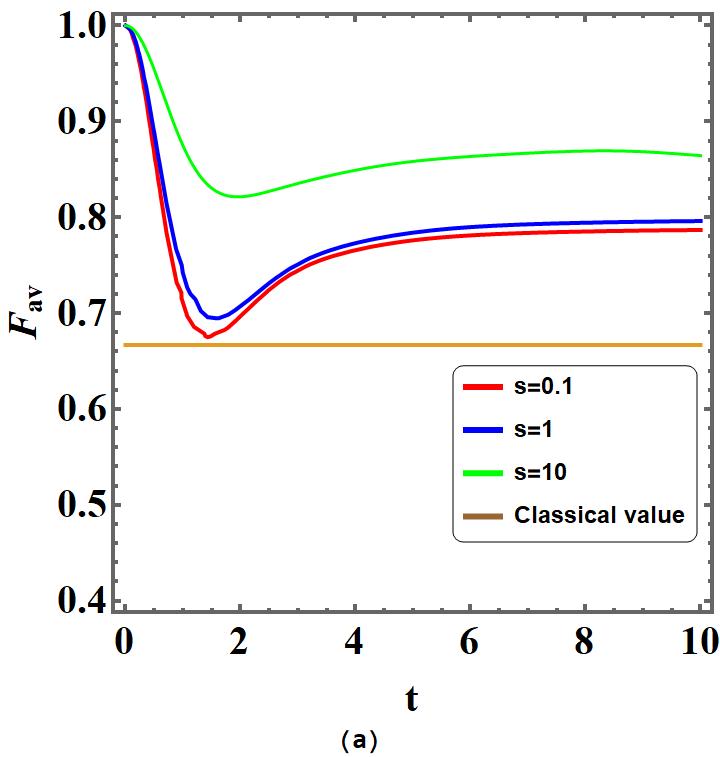} \hspace{2mm}
	\includegraphics[width=4cm,height=4cm]{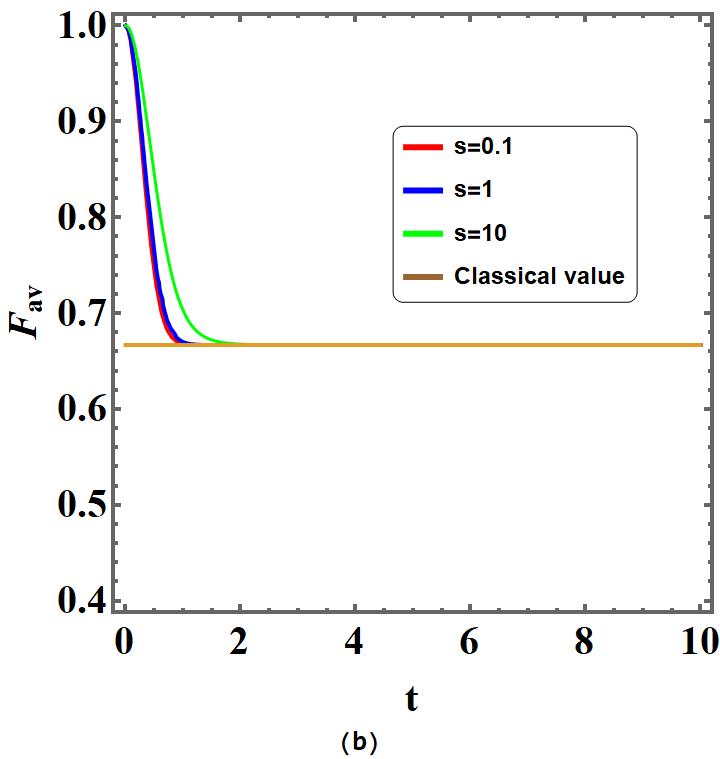} \hspace{2mm}
	\includegraphics[width=4cm,height=4cm]{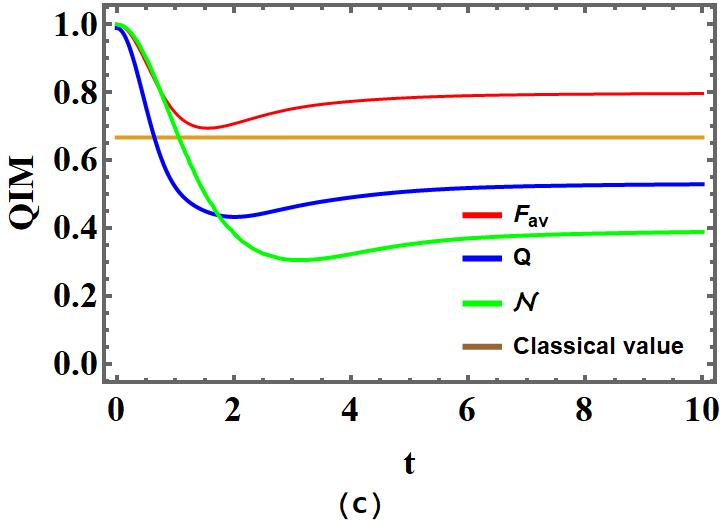} \hspace{2mm}
	\includegraphics[width=4cm,height=4cm]{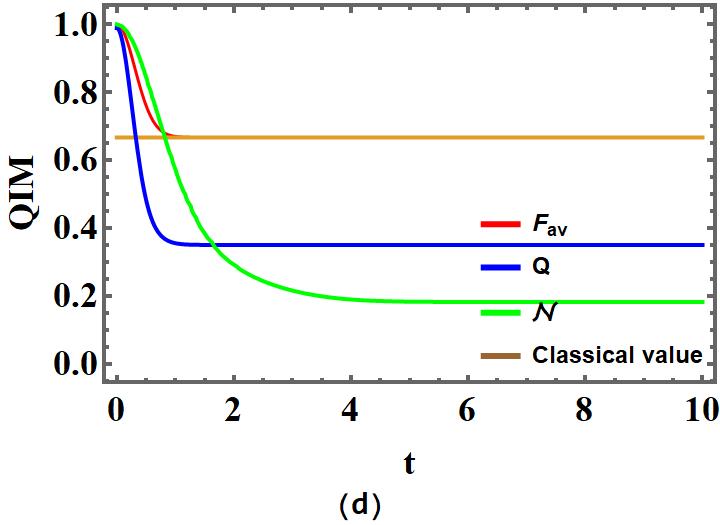}
	\caption{Variation of average fidelity $F_{av}$ with time fordifferent values of s at (a) low temperature $ \beta\omega_c \sim 1$ (b) high temperature $\beta \omega_c <<1$. Comparison of  quantum information measures ({\bf QIM}) given by $F_{av}$ , negativity $\mathcal{N}$ and discord $Q$ is made for (c) $ \beta\omega_c \sim 1$ and (d) $\beta \omega_c <<1$ for $s=1$.}
	\label{fig2}
\end{figure}
 \begin{eqnarray}
 \label{F1}
		F_{av} &=& \frac{1}{4\pi} \int_{0}^{\pi} d\theta\int_{0}^{2\pi} d\phi \bar{F} \sin\theta\\ \nonumber
		&=& \frac{2}{3} + \frac{ \kappa(t) + \kappa^*(t) }{6} = \frac{2}{3} + \frac{1}{3}\cos(2\zeta(t))e^{-\gamma(t)}.
\end{eqnarray}

 Since $\zeta(t)$ and $\gamma(t)$ depend on the parameters of the bath and distance of separation of qubits $L$, this  result therefore shows that average fidelity $F_{av}$ is independent of initial SB correlations. Next, we plot in figure \ref{fig2},  time dependence of $F_{av}$ for different values of $s$  for high temperature and low temperature cases. In the low temperature regime figure \ref{fig2}(a), we see that $F_{av}$ has strong non-Markovian behaviour, decaying first to a classical optimal value and then saturating at a value higher than classical value of $\frac{2}{3}$. Also, for large values of $s$, i.e. larger the qubit separation of channel, we see that $F_{av}$ is always greater than $\frac{2}{3}$. However, in the high temperature limit $\beta \omega_c<<1$, figure \ref{fig2}(b), we have $F_{av}$ saturating at a classical value for almost all values $s$.  As we increase the distance between the qubits of the channel, less number of modes interact that results in useful quantum correlations at $\beta \omega_c \sim 1$ (figure \ref{fig2}(c)). In the other case $\beta \omega_c <<1$, we see that thermal fluctuations play an important role to destroy long range correlations. Since $F_{av}$ does not depend on initial SB correlations which would compete with thermal fluctuations. Thus, due to small correlations  present in the $\beta \omega_c <<1$ case (figure \ref{fig2}(d)), we have $F_{av}$ saturating at classical value.  
  
As the second case, we consider the qubits with Alice are subjected to  decoherence in presence of initial SB correlations.  It can be shown that fidelity is also independent of initial SB correlations and is given by the same equation \ref{F1}.
 
In the third case, we consider the qubit that is being teleported subjected to decoherence. In this  case,  the average fidelity $F_{av}$ is dependent on initial SB correlations and is given by
 \begin{eqnarray}
     F_{av} = \frac{2}{3} + (\frac{\beta \omega_0}{2}-1)\frac{\cos(\zeta_0(t))}{6 \sinh \frac{\beta \omega_0}{2}}e^{-\gamma(t)} ,
 \end{eqnarray}
where $\zeta_0(t) = 4\sum_k \frac{|g_k|^2}{\omega_k^2 }\sin \omega_kt$. From this result, we observe that at the critical temperature $\frac{\beta \omega_0}{2} \sim 1$, we get classical bound of fidelity $F_{av} =\frac{2}{3}$. For $\beta \omega_0 <<1$, we have $F_{av} = \frac{2}{3}-\frac{\cos(\zeta_0(t))}{3 \beta \omega_0}e^{-\gamma(t)}<\frac{2}{3}$.  Therefore for high temperatures where thermal fluctuations destroy coherence and  we get $F_{av} <2/3$. In case of low temperatures $\beta \omega_0 >>1$, we have $F_{av}=\frac{2}{3}+\frac{\beta \omega_0}{6} e^{-\frac{\beta \omega_0}{2}} \cos(\zeta_0(t))e^{-\gamma(t)}$ which is always greater than $\frac{2}{3}$ for finite value of temperature.

\section{ Conclusion}
In conclusion, we have studied the role of initial SB correlations on the dynamics of quantum correlations given by entanglement and discord in  dephasing model with distance dependent interactions. The joint state of SB is constructed via projected measurements on an initially thermal equilibrium state of system and bath. In the low temperature regime $\beta \omega_c \sim 1$, we have shown that negativity and discord have non-monotonic behavior due  
to underlying non-Markovian effects present in the dynamics. Due to presence of initial SB correlations, we have negativity and discord saturating to a finite non-zero value. 

Next, in order to investigate the usefulness of these saturated values of quantum correlations in the long time limit, we studied the standard teleportation protocol. In the case, where channel is coupled to bath, we have shown that initial SB correlations have no role to play on average fidelity of teleportation. Moreover, we have shown the distance between the qubits of the channel effect the dynamics of average fidelity. In the low temperature case, the average fidelity is always greater than the classical value while for high temperature case, it saturates to classical value. 

The same results holds true if the qubits with Alice undergo dephasing dynamics. However, if the qubit that is being teleported is subjected to dephasing, the average fidelity strongly depends on the initial SB correlations. At high temperatures in this case, it is shown that due to  thermal fluctuations, the average fidelity is always less than classical value while at low temperatures it is saturates to classical value in the long time limit. Also, there exist a critical temperature $\frac{\beta \omega_0}{2}\sim 1$, for which $F_{av}\rightarrow \frac{2}{3}$.

 \end{document}